# Efficient anisotropic polariton lasing using molecular conformation and orientation in organic microcavities

*Florian Le Roux*[1*], *Andreas Mischok*[1], *Donal D. C. Bradley*[2], *Malte C. Gather*[1,3*]

[1] Humboldt Centre for Nano- and Biophotonics, Department of Chemistry, University of Cologne, Greinstr. 4-6, 50939 Köln, Germany

[2] Physical Science and Engineering Division, King Abdullah University of Science and Technology (KAUST), Thuwal 23955-6900 Saudi Arabia

[3] Organic Semiconductor Centre, SUPA School of Physics and Astronomy, University of St Andrews, St Andrews, KY16 9SS UK

**Keywords:** polariton lasing, exciton-polaritons, conjugated polymer microcavities, organic semiconductors, liquid crystalline conjugated polymers, molecular conformation and orientation.

**Abstract:** Organic exciton-photon polariton lasers have recently been shown to be remarkable candidates for the realization of efficient sources of coherent light operating at room temperature. While their thresholds are now comparable with conventional organic photon lasers, tuning of molecular conformation and orientation as a means to further enhance their performance remains largely unexplored. Here, we first report a two-fold reduction in the threshold of a polariton laser based on a high-Q microcavity filled with an active layer of poly(9,9-dioctylfluorene) (PFO) when 15% $\beta$-phase conformation is introduced. We then take advantage of the liquid crystalline properties of PFO and use a thin transparent sulfuric dye 1 (SD1) photoalignment layer to induce homogeneous nematic alignment of the polymer chains. The resulting transition dipole moment orientation increases the Rabi energy, bringing the system into the ultra-strong coupling regime where we observe anisotropic polariton lasing with an eight-fold reduction in absorbed threshold, down to 1.14 pJ / 0.36 μJcm$^{-2}$ for the direction parallel to the orientation, with no emission along the orthogonal direction. To our knowledge, this threshold is lower than demonstrated with state-of-the art optically pumped organic vertical cavity surface-emitting photon and polariton lasers. This demonstration opens new opportunities for more efficient polaritonic devices and the observation of fundamental effects at low polariton numbers.

# 1. Introduction

Exciton-polaritons are hybrid-states of light and matter whose energy signatures can be readily observed in planar microcavities when the interaction between the photonic and the semiconductor excitonic fields become larger than their respective losses. In turn, the separate energy characteristics of light and matter are no longer distinguishable.[1,2] Frenkel excitons found in organic semiconductors possess large binding energies and oscillator strengths[3,4] which, respectively, allow for room-temperature operation and a large Rabi splitting, $\hbar\Omega$, between the lower (LP) and upper (UP) polariton branches; values in excess of 1 eV have been demonstrated in organic microcavities.[5–8]

Organic polariton lasing, where the polaritons scatter to macroscopically occupy the ground state and then decay through the emission of coherent photons, has been observed more than a decade ago in a high Q-factor planar microcavity containing a single-crystal of anthracene.[9] Following this first demonstration, similar observations were reported in cavities containing layers of small organic molecules[10], polymers[11] or proteins[12], where the increased LP lifetime allowed for an efficient thermalization of the LP ground state population, enabling Bose Einstein Condensation (BEC). Since then, relying on similar structures, multiple fascinating demonstrations have been made, many of which make use of the quantum nature of polariton condensates, e.g. superfluidity of light,[13] optical logic at room-temperature,[14] and single-photon detection.[15] Considerable progress has been made in improving the performance of organic polariton lasers; the use of materials exhibiting high photoluminescence quantum yield (PLQY) and fast exciton decay rates has proven particularly beneficial in this context.[16,17] This enabled demonstrations of nonlinear thresholds in the range 2.2 µJ cm$^{-2}$ [16] to 9.7 µJ cm$^{-2}$,[17] values on par with the best organic photon lasers using a vertical cavity surface emitting design.[18,19] In terms of design, adjusting the active layer thickness to target the (0-1) vibronic transition in emission has been shown to be an effective way to further reduce the lasing threshold.[17]

Transition dipole moment orientation is a known method to reduce the threshold for optically pumped exciton gain in poly(9,9-dioctylfluorene) (PFO) waveguides.[20] It has also recently been suggested[21,22] as a way to further reduce the threshold of polariton lasers through efficient population of the LP ground-state.[22] Several material systems have been explored to realize 1D exciton-polaritons: carbon[23] and tungsten disulfide[24] nanotubes, organic single crystals[25] and liquid crystals[26] but polariton lasing from such systems remains a challenge. Thin, transparent photoalignment layers of an azobenzene-containing Sulfuric Dye (SD1) have recently been used to induce homogeneous nematic alignment of conjugated polymers in metallic microcavities,[27] leading to significantly enhanced Rabi splitting when the polarization of the incident light is aligned with the polymer transition dipole moment orientation direction. One of the polymers used in that work, PFO, has been successfully used to fabricate various electronic/optoelectronic devices (light-emitting diodes (LED)s including polarized light emission structures,[28,29] transistors,[30,31] and optically pumped photon lasers[32]). It can also be processed so as to adopt distinct conformations, allowing an exploration of physical structure control over the influence of exciton ensemble sub-populations on polariton formation and properties in metallic microcavities.[33] The disordered glassy-phase possesses an inter-monomer torsion angle centered around 135º, with a more or less Gaussian distribution of exciton energies.[34] Glassy PFO was previously used to demonstrate polariton lasing with an absorbed threshold pump energy density of 19.1 μJ cm$^{-2}$.[35] The linearly-extended β-phase conformation,[36,37] with a near-to-planar chain segment geometry, presents red-shifted absorption and emission spectra with characteristically-well-resolved vibronic progressions and an enhanced oscillator strength. It forms a distinct exciton sub-population located on the red edge of the glassy Gaussian distribution and the relative weighting of these populations allows polariton properties to be tuned.[33] The β-phase can be generated in thin films using several different methods, for example: (i) globally, via solvent vapour annealing,[37] use of solvent mixtures incorporating a high-boiling point additive,[38] and dipping in a moderate solvent or solvent/non-solvent mixture[37] and (ii) locally, via masking[36,39] or dip-pen patterning.[40] Interestingly, PL emission from a sample containing even a small fraction of β-phase is typically dominated by the spectral contribution of the β-phase due to rapid energy transfer from the glassy- to the β-phase chain segments.[37,41]

Here, we use PFO β-phase generation and nematic liquid crystal alignment to demonstrate an ability to tune the performance of polariton lasers via physical-structure-based, molecular level control of exciton properties. We first fabricate two high-Q microcavities made of distributed Bragg reflector (DBR) mirrors sandwiching spin-coated layers of PFO prepared in the glassy- (sample S1) and β-phase (sample S2 with 15% β-phase chain segments) conformations. Both cavities are designed so that the energy of the LP at normal incidence (θ = 0º) is tuned to 2.67 eV, which matches the (0-1) vibronic peak of the β-phase. We observe that targeting this energy level enables a reduction in the absorbed pump polariton lasing threshold for the glassy-phase cavity to 3.8 µJ cm$^{-2}$ (equivalent to 12.08 pJ per excitation pulse) from the 19.1 µJ cm$^{-2}$ value recorded in earlier reports.[35] Introducing 15% β-phase then allows for a further two-fold reduction in absorbed pump threshold to 1.8 µJ cm$^{-2}$ (5.71 pJ per pulse). In order to study the influence of the polymer chain alignment and corresponding transition dipole moment orientation, we then fabricate a high-Q microcavity containing a layer of PFO which is subjected to thermotropic alignment on a SD1 photoalignment layer, before the generation of 15% β-phase (sample S4). Orientation of the transition dipole moments in the active layer enables anisotropic coupling in which excitation with a pump polarization parallel to that orientation leads to an over eight-fold reduction in absorbed pump threshold from 2.9 µJ cm$^{-2}$ (9.14 pJ per excitation pulse) for an unaligned but otherwise comparable 15% β-phase cavity (sample S3), to 0.36 µJ cm$^{-2}$ (1.14 pJ per excitation pulse) for the aligned sample S4. The Rabi splitting energies for these structures are substantially enhanced in the direction parallel to the orientation, with the coupling strength $g_1 = \frac{\Omega_1}{\omega_G}$ reaching up to 21.7% and a consequent crossover into the ultra-strong coupling (USC) regime without the need to use metallic mirrors to squeeze the mode volume.[27,33]

## 2. Results and Discussion

**Figure 1 a)** shows the schematic structure of the samples used in this study. Sample S1 contains a 155 nm thickness glassy-phase PFO film, S2 a 152 nm thickness 15% β-phase film, S3 a 166 nm thickness 15% β-phase film, and S4 a 145 nm thickness 15% β-phase film aligned on top of a thin layer of SD1 (5 nm). In all samples, the active polymer layers are sandwiched between $Ta_2O_5/SiO_2$ alternating layer DBR mirrors. Samples 1 and 2 have an asymmetric cavity structure with 10.5 $Ta_2O_5/SiO_2$ pairs below and 7.5 above whilst samples 3 and 4 are symmetric cavity structures with 10.5 $Ta_2O_5/SiO_2$ pairs on both sides.

In order to remove any variations in the thickness of the mirror layers between samples as a possible source of error, we deposited the DBRs for pairs of samples at the same time, namely samples S1 and S2 together, and S3 and S4 together. The DBR mirrors for samples S1 and S2 are made of 50.7 nm thickness $Ta_2O_5$ / 74.4 nm thickness $SiO_2$ alternating layers whilst for samples S3 and S4 it is 48.8 nm $Ta_2O_5$ / 71.8 nm $SiO_2$. Samples S3 and S4 show a slight blue-shift (+0.07 eV) in the stop-band compared to samples S1 and S2 (see Figure S2). This shift leads to a small increase in reflection loss around the LP energy (2.66 - 2.67 eV). As a result, despite the increased number of layers in the top DBRs for S3 and S4, similar Q-factors of

~ 600 are found for all four samples (calculated from the LP emission linewidth at low excitation energy).

The polymer chains of spin-coated PFO films tend to lie within the plane of the film,[42] yielding a uniaxial in-plane/out-of-plane anisotropy in the corresponding optical constants. **Figure 1 b)** shows the ordinary components of the measured refractive indices ($n_{\text{ord}}$) and extinction coefficients ($k_{\text{ord}}$), obtained using variable angle spectroscopic ellipsometry (VASE) on films prepared as per samples S1 and S2 on glass substrates. The characteristic, inhomogenously broadened distribution corresponding to the $S_0 - S_1$ optical transition of the glassy-phase of PFO centered at around $X_G = 3.23$ eV is clearly visible. As reported in Ref. 37, the introduction of *β*-phase chain segments leads to the appearance of a clearly resolved (0-0) vibronic peak in absorption, which is centered at around $X_\beta$ = 2.86 eV. **Figure 1 c)** correspondingly shows the in-plane optical constants obtained for a PFO film thermally aligned on top of a SD1 photoalignment layer and then subjected to solvent vapour annealing to induce 15% *β*-phase. As expected, the oscillator strength is much stronger in the y-direction (parallel to the orientation of the polymer chains) than in the x-direction, confirming that the transition dipole moment is largely axial. The off-axis component arises from individual dipole moments lying at ~ 20-25° to the chain axis for glassy-phase PFO,[43,44] and reduces for the chain-extended *β*-phase conformation, leading to a higher anisotropy.[45] **Figure 1 d)** shows PL spectra for non-aligned glassy- and 15% *β*-phase PFO films. As previously reported,[37, 41] the spectra differ substantially: for the glassy-phase film the vibronic peaks appear at 2.93 eV (0-0), 2.75 eV (0-1), and 2.57 eV (0-2) whereas for the *β*-phase film the peaks are at 2.82 eV (0-0), 2.67 eV (0-1), and 2.50 eV (0-2).

The deduced refractive indices were used in a transfer matrix calculation to determine the PFO film thicknesses for each sample such that at normal incidence (i.e., *q* = 0) the lower polariton energy, $E_{LP}$, matches the (0-1) emission vibronic peak of the β-phase at 2.67 eV.

**Figure 1**. a) Schematic of the different samples (S1, S2, S3, S4) used in this study as well as the chemical structures of the glassy- and β-phase chain conformations of PFO. b) In-plane optical constants for glassy-phase (blue lines) and 15% β-phase (red lines) PFO thin films showing the ordinary parts of the extinction coefficient (solid lines), $k_{\mathrm{ord}}$, and refractive index (dashed lines), $n_{\mathrm{ord}}$. c) The in-plane, x- (black lines) and y- (red lines), parts of the extinction coefficient, $k_x$ and $k_y$ (solid lines), and refractive index, $n_x$ and $n_y$ (dashed lines) for a PFO thin film aligned along the y-direction and subjected to 15% β-phase generation. d) PL spectra for spin-coated PFO thin films in the glassy phase (blue line) and containing 15% β-phase (red line).

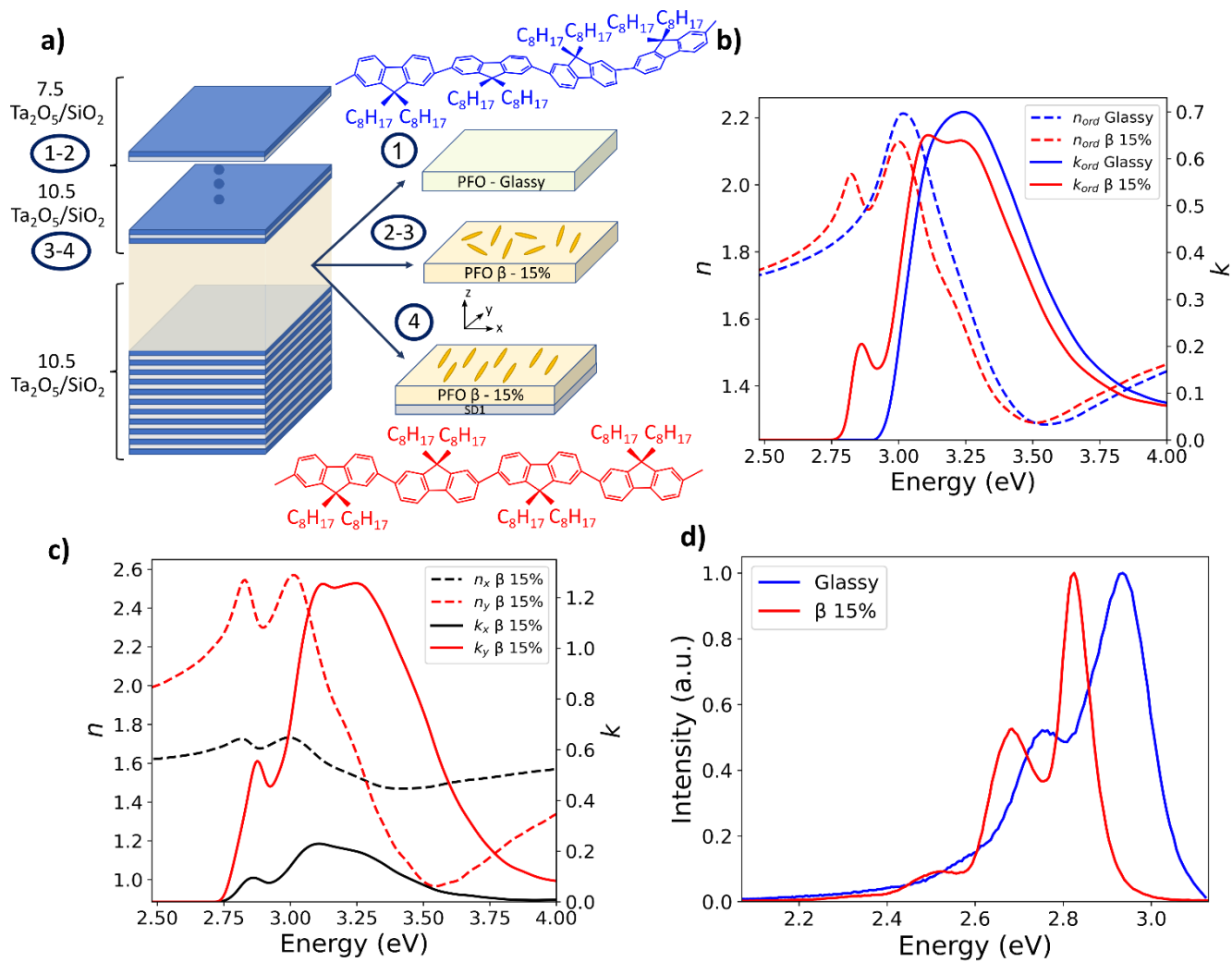

Transverse electric (TE) polarized white light reflectivity spectra were simulated for each sample and for a range of angles of incidence using transfer matrix calculations as shown in **Figure 2**. The simulated spectra are in good agreement with measurements of DBR transmissivity (shown in Figure S2) and with polymer thickness measurements using a profilometer. For sample S4, the transmissivity was simulated for an alignment of the polymer chains parallel to the polarization of the incident light (i.e., the azimuthal angle $\varphi$ between the TE-polarization and the alignment direction was 0°). The LP energy was fitted using the Hopfield-Agranovich model[46,47] with one excitonic transition for the glassy phase sample S1 ($X_G$) and two transitions for the $\beta$-phase samples S2, S3, and S4 ($X_G$ and $X_\beta$). The presence of two Rabi splitting energies, $\hbar\Omega_1$ for $X_G$ and $\hbar\Omega_2$ for $X_\beta$, indicates that both conformations contribute as discussed in more detail in Ref. 21. The results from the fitting procedure are summarized in **Table 1**.

**Figure 2 a)** shows the optical characterization for S1. The Hopfield-Agranovich fitting yields a Rabi splitting energy of $\hbar\Omega_{1,S1} = 0.52$ eV, which is in good agreement with previous reports.[35] The detuning $\Delta_{1G}$ = - 465 meV between $X_G$ and the energy of the photonic mode at normal incidence $E_{ph}(\theta = 0°) = \hbar\omega_{cav,q=0}$ was chosen to be significantly larger than in previous reports in order to target the favorable (0-1) transition. **Figure 2 b)** shows the optical characterization for S2, where two detuning energies exist: first, $\Delta_{2G} = -470$ meV similar for the same reason to $\Delta_{1G}$ for S1, and second, $\Delta_{2\beta} = -100$ meV between $X_\beta$ and $E_{ph}(\theta = 0°)$. The ratio of the Rabi energies $\hbar\Omega_{1,S2} = 0.50$ eV and $\hbar\Omega_{2,S2} = 0.05$ eV is similar to that found in Ref. 35, where it was shown that the larger oscillator strength of $\beta$- relative to glassy-phase excitons and the interplay of their vibronic structures is important in understanding these coupling strengths. The detuning energies and Rabi splitting values for S3 in **Figure 2 c)** are, as expected, similar to those for S2. Notable differences can, however, be found in **Figure 2 d)**, where the alignment of PFO chains for S4 increases the Rabi splittings to $\hbar\Omega_{1,S2} = 0.70$ eV and $\hbar\Omega_{2,S2} = 0.07$ eV. It was previously observed[26,27] that the increase in Rabi splitting following alignment of the molecular units is due to the increase of the dot product, $\boldsymbol{\mu} \cdot \boldsymbol{E}$, between the molecular transition dipole moment vector, $\boldsymbol{\mu}$, and the incident electric field vector, $\boldsymbol{E}$. Following alignment, the resulting coupling strength $g = \frac{\hbar\Omega_{1,S4}}{\hbar\omega_G}$ reaches 21.7% which brings the system into the USC regime. Previous demonstrations of USC in organic materials have frequently required a cavity formed by metallic mirrors to reduce the mode volume.[5,33]

Figure 2. Transfer matrix simulations of angle-resolved, TE-polarized reflectivity spectra for a) sample S1 containing non-aligned glassy-phase PFO, b) and c) samples S2 and S3 containing non-aligned 15% *β*-phase PFO, and d) sample S4 containing aligned 15% *β*-phase PFO. The horizontal white lines indicate the position of the excitonic transitions $X_G$ and $X_β$ as described in the text. The curved white line shows the fitted cavity mode $E_{ph}(θ) = ℏω_{cav,q}$ and the dashed orange line shows the fitted LP mode. For S4, the simulation was performed at $φ=0°$, i.e. PFO polymer chains parallel to polarization of incident light.

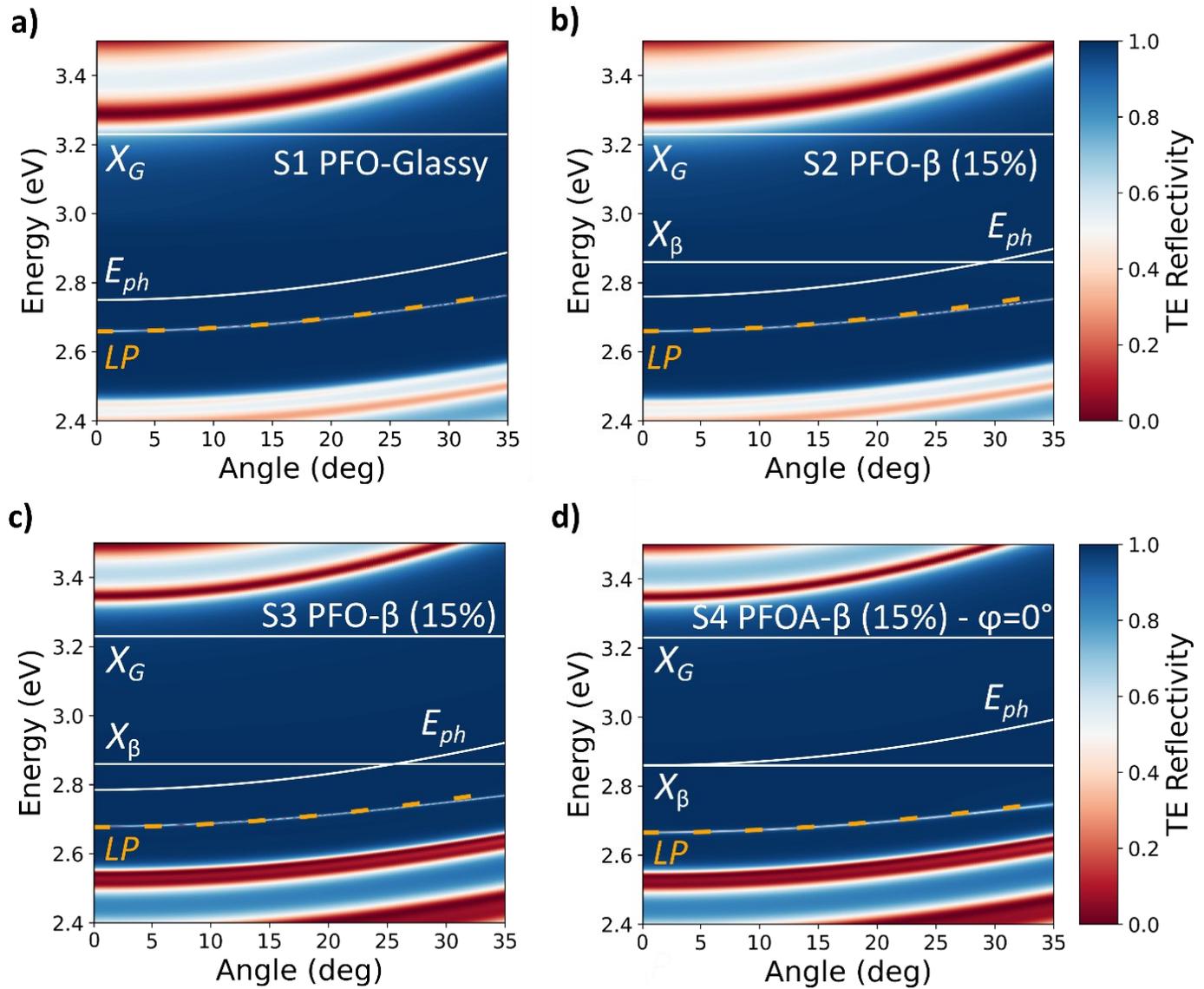

**Table 1.** Extracted and preset parameter values for microcavities S1, S2, S3 and S4, modelled using a Hopfield-Agranovich Hamiltonian.[46,47]

| Sample | $\hbar\omega_{cav}(\theta = 0°)$ [eV] [a] | $\hbar\omega_G$ [eV] [b] | $\hbar\omega_\beta$ [eV] [c] | $n_{eff}$ [d] | $\hbar\Omega_1$ [eV] [e] | $\hbar\Omega_2$ [eV] [f] |
|---|---|---|---|---|---|---|
| S1 (Glassy-phase with 7.5 pair top DBR) | 2.765 | 3.23 | – | 1.88 | 0.52 | – |
| S2 (15% β-phase with 7.5 pair top DBR) | 2.76 | 3.23 | 2.86 | 1.88 | 0.50 | 0.050 |
| S3 (15% β-phase with 10.5 pair top DBR) | 2.785 | 3.23 | 2.86 | 1.90 | 0.51 | 0.053 |
| S4 (Aligned 15% β-phase with 10.5 pair top DBR) | 2.86 | 3.23 | 2.86 | 1.95 | 0.70 | 0.070 |

[a] Fitted energy of the bare cavity mode at normal incidence

[b] Preset exciton oscillator transition energy at the peak of the vibronically-unresolved ensemble transition (S$_0$-S$_1$) of glassy-phase PFO films

[c] Preset exciton oscillator transition energy at the center of the (0-0) vibronic absorption peak of β-phase PFO films

[d] Fitted microcavity effective refractive index

[e] Fitted Rabi splitting energy associated with transition $\hbar\omega_G$

[f] Fitted Rabi splitting energy associated with transition $\hbar\omega_\beta$

Next, the angle-resolved emission from the microcavities was measured via Fourier plane imaging. The samples were excited non-resonantly at 355 nm (3.49 eV), using 25 ps pulses from a diode-pumped Nd-YAG laser at a repetition rate of 250 Hz. **Figure 3 a)** shows the emission from cavity S1 both at low and at high pump pulse energies. At low pulse energies, the emission is spectrally broad and closely follows the fitted LP shown in Figure 2 a). Above a certain threshold, however, the emission dramatically reduces in linewidth, blue-shifts, and collapses to a small angle range around the normal to the plane. Similar observations are made for S2 in **Figure 3 b)** and S4 at $\varphi=0°$ in **Figure 3 c)**. **Figure 3 d)** shows the emission for S4 at $\varphi=90°$, where the excitation is perpendicular to the polymer chain alignment. The background-to-emission ratio is higher in Figure 3 c) and Figure 3 d) compared with Figure 3 a) and Figure 3 b) as the overall excitation power was lower due to the decreased threshold for S4 relative to S1 and S2. No emission is recorded from the orthogonal alignment direction due to the strong anisotropy of the system. We also note that in each case, the emission acquires linear polarization parallel to the polarization of the excitation, and, therefore, for S4 also parallel to the chain alignment. When the pump excitation polarization is gradually moved away from the chain alignment direction, it is still possible to observe lasing but with an increasing excitation threshold due to the reducing absorption. The linear polarization acquired by the laser emission from the sample remains, however, largely parallel to the main component of the transition dipole moment since the 'chromophores' that absorb light will emit along the transition dipole moment direction.

**Figure 3.** Angle-resolved PL spectra intensity maps for microcavities a) S1, b) S2, c) S4 at $\varphi=0°$ and d) S4 at $\varphi=90°$ (no emission visible) with emission spectra below (left half) and above (right half) threshold. For each panel, two intensity scales are used.

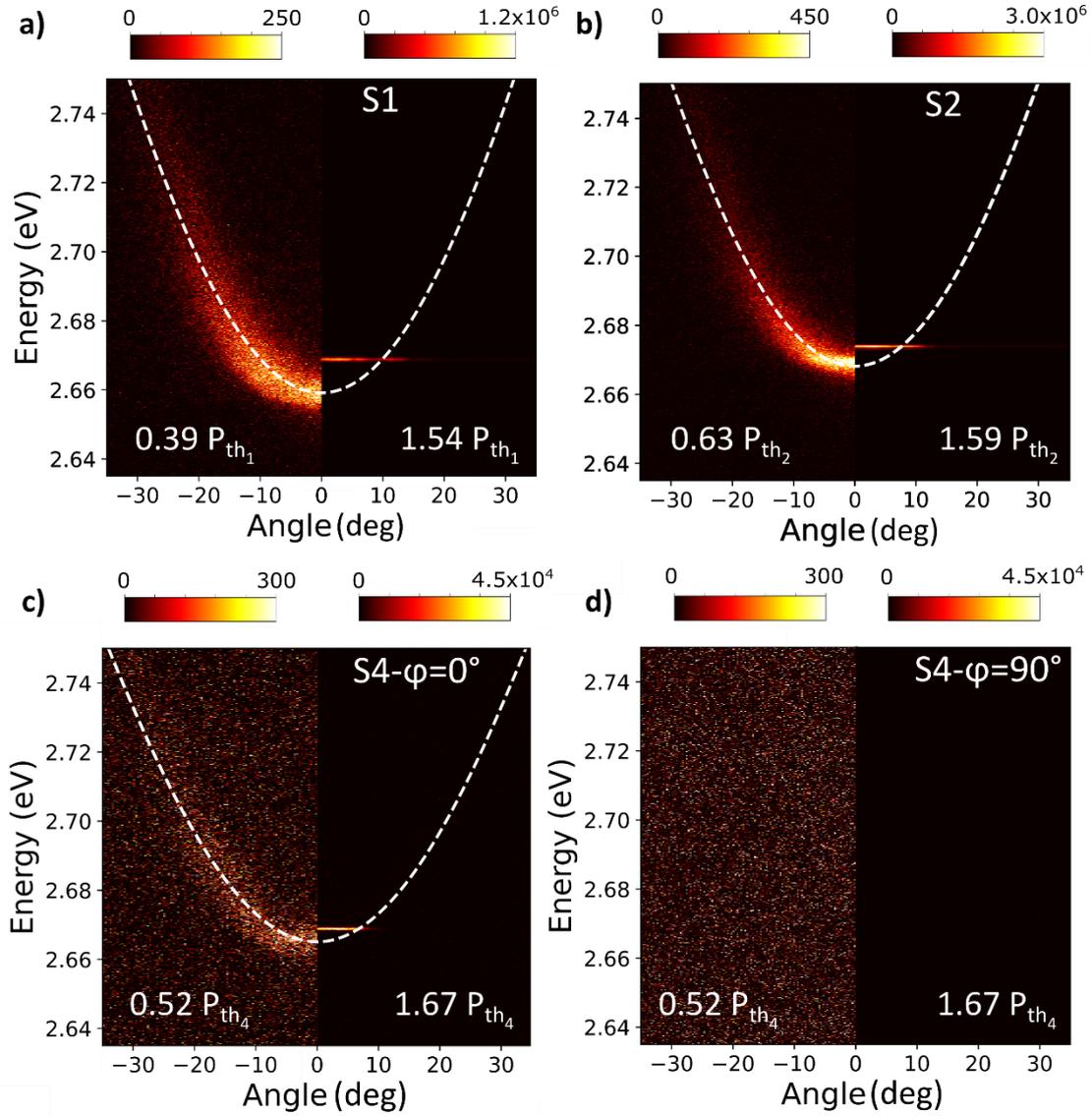

**Figures 4 a)**, and **c)** display for all samples the emission intensity for increasing incident excitation pulse energies, integrated in each case over $\theta \in [-2°; 2°]$ and $E \in [2.635 \text{ eV}; 2.75 \text{ eV}]$. The full PL spectra used for these integrations are shown in Supporting Information Figure S1. For each sample, a clear super-linear increase in intensity is observed above a specific threshold excitation energy $P_{\text{th}}$. The saturation in emission intensity observed for S1 and S2 above threshold in Figure 4 a) was also observed in Ref 35. Comparing the pair S1 and S2, we observe that the introduction of 15% β-phase chain segments in sample S2 leads to a two-fold reduction in threshold pulse energy relative to S1, from $P_{\text{th}_1} = 25.16$ pJ down to $P_{\text{th}_2} = 12.41$ pJ. We attribute this reduction in threshold to the faster radiative decay of β-phase excitons[37, 48] and the increased overlap of the LP at normal incidence with the (0-1) emission peak of the β-phase.[17]

Introducing small fractions of β-phase into glassy PFO films is known to increase the PLQY from ~55% for glassy PFO films up to almost 70% in an optimal case that can be reached through the use of alkyl additives in the initial solution or with a solvent/non-solvent dipping method.[42] However, for the relatively large fraction of β-phase present in S2, the PLQY is expected to be similar to the glassy PFO of S1,[41] indicating that the reduction in threshold for S2 is most likely not the result of increased PLQY.

In Figure 4 c), we observe that the orientation of the transition dipole moments leads to an almost 7-fold reduction in threshold, from $P_{\text{th}_3} = 14.50$ pJ for S3 down to $P_{\text{th}_4} = 2.23$ pJ for S4. We attribute this reduction to the efficient pumping of the exciton reservoir as the excitation pump polarization is now aligned with the main component of the transition dipole moment. In turn, nearly all generated excitons can efficiently take part in the condensation process. It is important to note that this reduction is achieved despite additional parasitic absorption introduced by the SD1 layer (Supporting Information Figure S3). $P_{\text{th}_3}$ is slightly larger than $P_{\text{th}_2}$ due to the slight blue-shift of the DBR stop-band in S3 and S4 compared to samples S1 and S2 (Supporting Information Figure S2).

Figure 4 b), d) display the linewidth and the spectral shift of the emission peak *versus* the incident excitation energy per pulse. For each sample we observe a drop in linewidth around the threshold value, corresponding to the increase in coherence acquired as the ground state of the LP is macroscopically populated.[10,11] As the excitation energy increases, we also observe a blue-shift of the emission peak caused by polariton-polariton and polariton-exciton interactions following depletion of the ground state.[10,49]

Combined, the reported observations clearly indicate the presence of polariton lasing for all four samples studied here. Spatial coherence measurements performed on the emission using a retro-reflector configuration Michelson interferometer allows independent corroboration of this conclusion. Above threshold, interference fringes with high visibility are observed (Supporting Information Figure S3), confirming the presence of spatial coherence and supporting the existence of polariton lasing.

**Table 2** presents a summary of the laser performance parameters for all four samples, comparing the incident and the absorbed pump pulse energy at threshold as well as the corresponding values of threshold pump fluence. To compute the absorbed pulse energy and fluence, the amount of the 355-nm pump light that is absorbed by the PFO film is calculated using transfer matrix calculations[50] (Supporting Information Figure S2). The excitation pump spot is 20 µm in our experiments. The absorbed threshold fluences are: $F_{th_1} = 3.8$ µJ cm$^{-2}$, $F_{th_2} = 1.8$ µJ cm$^{-2}$, $F_{th_3} = 2.9$ µJ cm$^{-2}$ and $F_{th_4} = 0.36$ µJ cm$^{-2}$. We attribute the reduction in $F_{th_1}$, relative to an earlier report for a similar microcavity,[35] to the difference in detuning between the two experiments (~150 meV), with the experiment presented here bringing the LP energy at normal incidence close to the (0-1) vibronic peak of glassy-phase PFO. The introduction of β-phase PFO and the preferential orientation of the transition dipoles leads to further dramatic reductions in threshold energy and threshold fluence. To our knowledge, the threshold absorbed fluence for polariton lasing in the aligned polymer microcavity, $F_{th_4} = 0.36$ µJ cm$^{-2}$, represents the lowest threshold observed to date for organic vertical cavity surface emitting polariton and photon lasers. In addition, the absolute absorbed pulse energy at threshold of 1.14 pJ is possibly the lowest among all organic semiconductor lasers reported so far. Whether pulse fluence or pulse energy is the more important figure of merit depends on the intended application of the laser.

**Figure 4. a)** and **c)** Integrated emission intensity versus incident excitation pulse energy for a) S1 (blue symbols) and S2 (black symbols) and c) S3 (black symbols) and S4 (red symbols). The polariton lasing thresholds are determined from the intersects of the fitted dashed lines and are listed next to each data set. **b)** and **d)** peak blue-shift (dashed line) and LP linewidth (solid line) for b) S1 (blue) and S2 (black) and d) S3 (black) and S4 (red). Vertical dashed lines indicate the polariton lasing thresholds as determined in a) and c). For S4, the measurements were performed at $\varphi = 0°$.

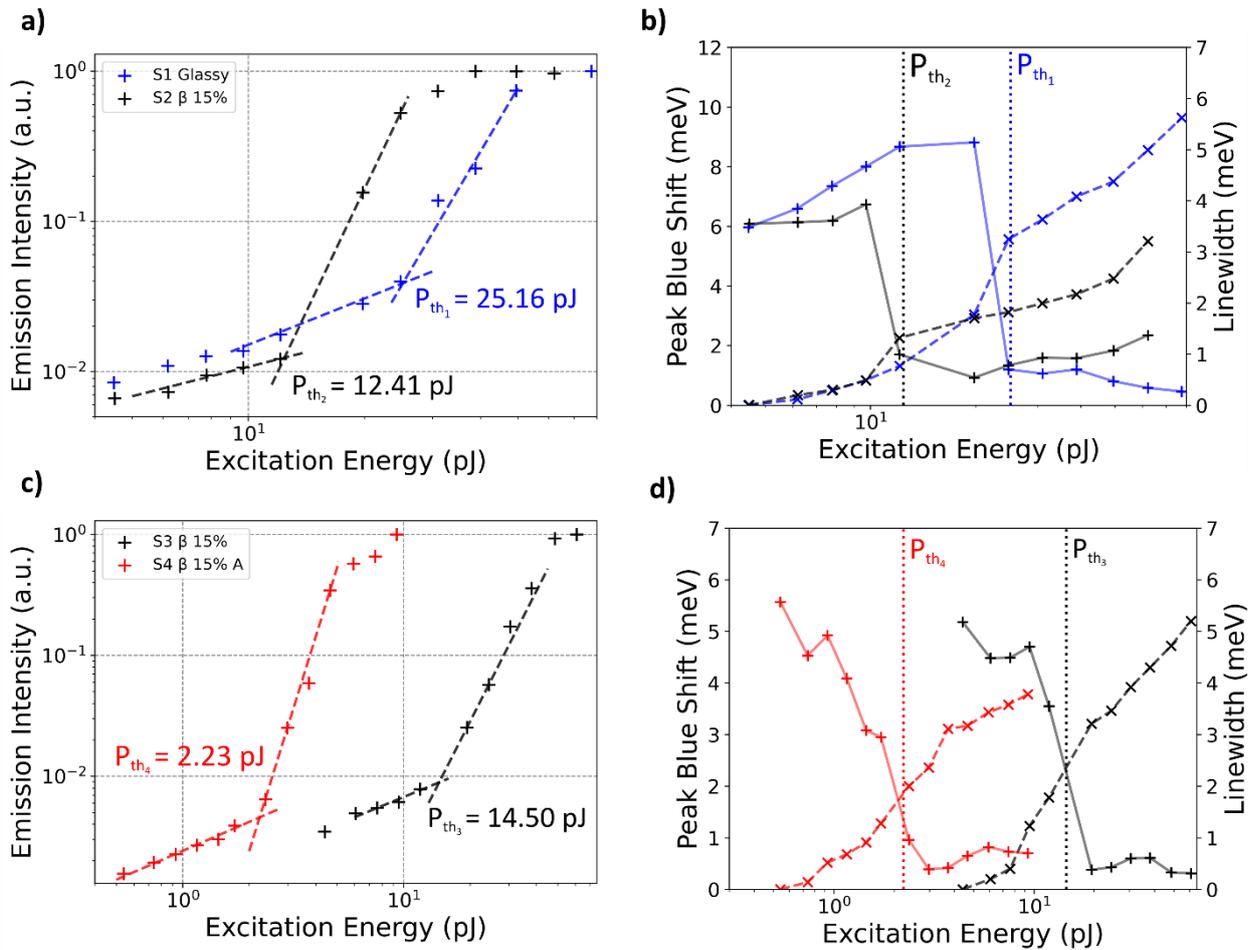

**Table 2.** Parameters extracted for polariton lasing thresholds of samples S1, S2, S3 and S4.

| Sample | $E_{LP}$ ($\theta = 0°$) Sub-threshold [eV] [a] | Incident Threshold Pulse Energy $P_{th}$ [pJ] | Intracavity PFO Absorption at 355 nm | Absorbed Threshold Pulse Energy $P_{th}$ [pJ] | Threshold Incident Fluence [µJ cm$^{-2}$] | Threshold Absorbed Fluence $F_{th}$ [µJ cm$^{-2}$] |
|---|---|---|---|---|---|---|
| S1 (Glassy / 7.5 pairs top DBR) | 2.66 | 25.16 | 48% | 12.08 | 8.0 | 3.8 |
| S2 ($\beta_{15\%}$ / 7.5 pairs top DBR) | 2.66 | 12.41 | 46% | 5.71 | 4.0 | 1.8 |
| S3 ($\beta_{15\%}$ / 10.5 pairs top DBR) | 2.67 | 14.50 | 63% | 9.14 | 4.6 | 2.9 |
| S4 (Aligned, $\beta_{15\%}$ / 10.5 pairs top DBR) | 2.66 | 2.23 | 51% | **1.14** | 0.71 | **0.36** |

[a] Energy of LP at normal incidence

## 3. Conclusion

We have demonstrated how the control of polymer conformation and morphology (specifically chain alignment), together with targeted spectral detuning can be used to improve the performance of PFO organic polariton lasers. First the use of a conformation change from glassy- to 15% $\beta$-phase, together with targeting the corresponding (0-1) vibronic emission peak, halved the absorbed polariton lasing threshold pulse energy and fluence to 5.71 pJ and 1.8 µJ cm$^{-2}$, respectively. The faster radiative decay for $\beta$-phase chain segments is considered to play a key role in this. We then demonstrated that in-plane nematic orientation of the polymer chain transition dipole moments inside the cavity leads to an impressive (more than 8-fold) reduction in absorbed threshold from 9.14 pJ and 2.9 µJ cm$^{-2}$ to 1.14 pJ and 0.36 µJ cm$^{-2}$. Compared to the original, glassy-phase device, the lasing threshold is reduced over 10-fold, with a further reduction expected if the stop band of the DBRs is fully optimized in the sample with nematic orientation. In addition, it is anticipated that a further increase in the PLQY of PFO through the use of a lower fraction of *β*-phase[37] combined with the use of additional DBR layers should allow a further reduction in threshold. Spatial patterning of the SD1 alignment layer using either a photomask[51] or direct two-photon laser writing[52] represents an additional exciting avenue that may allow further improvement in the performance of polariton lasers through in-plane confinement of polaritons and, additionally, offers the opportunity to study topologic phenomena[53] without a need to pattern the cavity mirrors.

## 4. Experimental Section/Methods

*Materials*: PFO was supplied by the Sumitomo Chemical Company, Japan and used as received. The peak molecular weight was $M_{pPFO} = 50 \times 10^3 \text{g mol}^{-1}$. D1 was supplied by Dai-Nippon Ink and Chemicals, Japan. Anhydrous toluene (99.85%) and anhydrous 2-methoxyethanol (≥ 99.8%) were purchased from Sigma-Aldrich and used as received. For the microcavity mirrors, $Ta_2O_5$ and $SiO_2$ were sputtered from >99.99% oxide targets (Angstrom Engineering). The substrates used were display-grade glass (Eagle XG, Howard Glass), 24 mm x 24 mm.

*Film fabrication:* The bare films used for ellipsometry in Figure 1 b) and c) and PL in Figure 1 d) were spin-coated from 18 mg mL$^{-1}$ PFO solutions in toluene in an inert environment. The solution was prepared in an inert environment (controlled atmosphere nitrogen glovebox with solvent filter) and left to stir overnight at a temperature of 50 °C. It was then filtered using a 0.45 μm PTFE filter. For the sample containing aligned chains of PFO, reported in Figure 1c), a layer of SD1 was first spin-coated (5s at 500 rpm, followed by 25s at 2000 rpm) from a 1 mg mL$^{-1}$ solution in 2-methoxyethanol and was then annealed for 10 min at 150 °C to evaporate any traces of solvent. Preparation of the SD1 as an alignment layer was performed in air by exposing the sample to polarized UV light (generated by a UV LED and a broadband polarizer, M365LP1 and WP25M-UB, Thorlabs) at a power of 5 mW for 10 min. A layer of PFO was then spin-coated on top of the SD1 in an inert environment, using 18 mg mL$^{-1}$ PFO in toluene solution, for 1 min at a speed of 2000 rpm, with an initial acceleration of 1000 rpm s$^{-1}$. Next, the sample was placed on a precision hotplate (Präzitherm, Gestigkeit GmbH) in an inert environment and the temperature was raised from 25 °C to 160 °C at a rate of approximately 30 °C min$^{-1}$. The upper temperature was then held for 10 min, followed by rapid quenching to room temperature by placing the sample on a metallic surface. For three (including the aligned sample) of the four film samples, an approximately 15% *β*-phase fraction was induced by exposing the films to a saturated toluene vapour environment for 24 hours. The thicknesses of the films and thus of the active layers in the final cavity were measured using a profilometer (Dektak) on simultaneously prepared reference samples and later confirmed by comparing the spectroscopic characterization to transfer matrix calculations.

*Microcavity Fabrication:* The microcavities were fabricated by radiofrequency magnetron sputtering of alternating layers of $SiO_2$ and $Ta_2O_5$ at a base pressure of $10^{-7}$ Torr, using 18 standard cubic centimeters per minute (sccm) Argon flow at 2 mTorr process pressure and 18 sccm Argon together with 4 sccm Oxygen flow at 4 mTorr process pressure for $SiO_2$ and $Ta_2O_5$, respectively. The additional oxygen flow during $Ta_2O_5$ deposition prevents the formation of unwanted sub-oxides. Spin-coating of the active layers was performed on top of the bottom mirror using the same process described in the *Film Fabrication* paragraph above, except that the concentration of the PFO solution was varied according to the desired film thickness: for S1 and S2, 27 mg mL$^{-1}$ PFO in toluene was used; for S3, 28 mg mL$^{-1}$ was used and for S4, 24 mg mL$^{-1}$. The SD1 alignment layer preparation process for sample S4 followed the description in the *Film Fabrication* section above.

*Characterization:* Time-integrated photoluminescence measurements are taken using a fluorescence spectrometer (Picoquant, FluoTime 250) under excitation by a picosecond pulsed laser operating at 375 nm (Picoquant, P-C-375). The permittivity of the different films is determined by variable angle spectroscopic ellipsometry measurements (VASE, M2000, J.A. Woollam) and subsequent modelling (via CompleteEase software, J.A. Woollam) with a uniaxial anisotropic B-spline for S1, S2, S3 and a biaxial anisotropic B-spline for S4. The transmissivity at normal incidence is determined using the same ellipsometer.

*Polariton Modelling:* The minima of the reflectivity maps shown in Figure 2 were analyzed using a least-squares fitting algorithm for the eigenvalue problem:

$$\boldsymbol{H_q v_{i,q} = \omega_{i,q} v_{i,q}} \tag{S1}$$

$$H_q = \begin{pmatrix} \omega_{cav,q} + 2D_q & -i\frac{\Omega_{1,q}}{2} & -i\frac{\Omega_{2,q}}{2} & -2D_q & -i\frac{\Omega_{1,q}}{2} & -i\frac{\Omega_{2,q}}{2} \\ i\frac{\Omega_{1,q}}{2} & \omega_G & 0 & -i\frac{\Omega_{1,q}}{2} & 0 & 0 \\ i\frac{\Omega_{2,q}}{2} & 0 & \omega_\beta & -i\frac{\Omega_{2,q}}{2} & 0 & 0 \\ 2D_q & -i\frac{\Omega_{1,q}}{2} & -i\frac{\Omega_{2,q}}{2} & -\omega_{cav,q} - 2D_q & -i\frac{\Omega_{1,q}}{2} & -i\frac{\Omega_{2,q}}{2} \\ -i\frac{\Omega_{1,q}}{2} & 0 & 0 & i\frac{\Omega_{1,q}}{2} & -\omega_G & 0 \\ -i\frac{\Omega_{2,q}}{2} & 0 & 0 & i\frac{\Omega_{2,q}}{2} & 0 & -\omega_\beta \end{pmatrix} \quad (S2)$$

In the case of a single exciton oscillator, $H_q$ reduces to the usual 4x4 Hopfield-like USC matrix[48,49]. $q$ is the in-plane wave vector, $\omega_{cav,q}$ the cavity mode energy, $\omega_j$ the frequency for the $j^{th}$-excitons ($j \in \{G,\beta\}$), $\Omega_{j,q}$ the associated Rabi frequency, for a given angle $\theta$, $\Omega_{j,q}(\theta) = \Omega_{0,j}\sqrt{\frac{\omega_j}{\omega_{cav}(\theta)}}$, where $\Omega_{0,j}$ is the Rabi frequency on resonance for the $j^{th}$-transition. In dielectric cavities, $\omega_{cav(TE),q} = \omega_{cav(TE)}(\theta) = \omega_{cav}(0)\left(1 - \frac{\sin^2(\theta)}{n_{eff}^2}\right)$. $D_q = \sum_j \frac{\Omega_{j,q}^2}{4\omega_j}$ is the contribution of the squared magnetic vector potential. The rest of the fitting procedure is identical to that described in Ref. 27.

*Angle-Resolved PL Measurements:* PL spectra were measured using Fourier imaging spectroscopy by imaging the back focal plane of a Nikon Plan Fluor (magnification 40×, numerical aperture 0.75) objective, set up on a Nikon Tie2-Eclipse microscope in reflection configuration. The THG output of a diode-pumped Nd-YAG laser (PL2210A, Ekspla), wavelength 355 nm, repetition rate 250 Hz, pulse duration 25 ps, was used for the excitation. Linear polarization of the pump was set by placing a Glan-Taylor polarizer (GT10-A, Thorlabs) before the objective. The diameter of the Gaussian beam at the sample plane was measured to be around 20 μm. The emitted light was directed towards the entrance of a spectrograph (Shamrock SR-500i-D2-SiL, Andor) equipped with an 1800 lines mm$^{-1}$ grating blazed at 500 nm and the PL spectra were imaged on an EM CCD camera (Newton 971, Andor) providing a spectral resolution of 40 pm. The spatial coherence measurements presented in the Supporting Information were performed using a Michelson interferometer in the retro-reflector configuration and the resulting interferograms were imaged on an sCMOS camera (ORCA-Flash 4.0, Hamamatsu).

**Supporting Information**

Supporting Information is available from the Wiley Online Library or from the author.

**Acknowledgements**

F.L.R acknowledges funding from the Alexander von Humboldt Foundation through a Humboldt Fellowship. A.M. acknowledges funding from the European Union Horizon 2020 research and innovation programme under the Marie Skłodowska-Curie grant agreement No. 101023743 (PolDev). D.D.C.B. thanks the Sumitomo Chemical Company and DIC Corporation for provision of PFO and SD1, respectively. D.D.C.B. further acknowledges support from King Abdullah University of Science and Technology, Jiangsu Industrial Technology Research Institute (JITRI) and JITRI-Oxford IMPACT Institute (R57149/CN001). This research was financially supported by the Alexander von Humboldt Foundation (Humboldt Professorship to M.C.G.) and by the European Research Council under the European Union Horizon 2020 Framework Programme (FP/2014-2020)/ERC Grant Agreement No. 640012 (ABLASE).

Received: ((will be filled in by the editorial staff))
Revised: ((will be filled in by the editorial staff))
Published online: ((will be filled in by the editorial staff))

# Efficient anisotropic polariton lasing using molecular conformation and orientation in organic microcavities

*Florian Le Roux[1]\*, Andreas Mischok[1], Donal D. C. Bradley[2], Malte C. Gather[1,3]\**

[1] Humboldt Centre for Nano- and Biophotonics, Department of Chemistry, University of Cologne, Greinstr. 4-6, 50939 Köln, Germany

[2] Physical Science and Engineering Division, King Abdullah University of Science and Technology (KAUST), Thuwal 23955-6900 Saudi Arabia

[3] Organic Semiconductor Centre, SUPA School of Physics and Astronomy, University of St Andrews, St Andrews, KY16 9SS UK

**Table of Contents**

Aligning poly(9,9-dioctylfluorene) (PFO) as active layer on top of a thin sulfuric dye 1 (SD1) photoalignment layer and then generating 15% *β*-phase conformation enables observation of anisotropic polariton lasing from high-Q factor microcavities, with an absolute threshold energy outperforming state-of-the art optically pumped organic lasers. This work highlights the importance of optimizing molecular physical structure for a new generation of efficient organic laser devices.

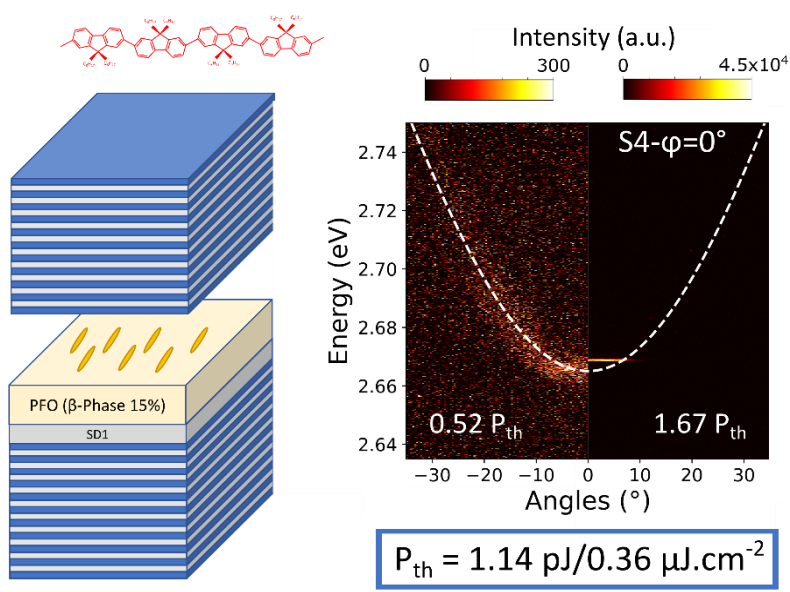

**TOC figure** – 55 mm broad x 50 mm high

# Supporting Information

**Efficient anisotropic polariton lasing using molecular conformation and orientation in organic microcavities**

*Florian Le Roux[1]\*, Andreas Mischok[1], Donal D. C. Bradley[2], Malte C. Gather[1,3]\**

[1] Humboldt Centre for Nano- and Biophotonics, Department of Chemistry, University of Cologne, Greinstr. 4-6, 50939 Köln, Germany

[2] Physical Science and Engineering Division, King Abdullah University of Science and Technology (KAUST), Thuwal 23955-6900 Saudi Arabia

[3] Organic Semiconductor Centre, SUPA School of Physics and Astronomy, University of St Andrews, St Andrews, KY16 9SS UK

**Figure S1**. Emission spectra with increasing excitation energy, transitioning from PL below threshold to lasing above. Data were collected over an angular range $\theta \in [-2°; 2°]$ for samples a) S1, glassy-phase PFO, 7.5 pairs top DBR, b) S2, 15% β-phase PFO, 7.5 pairs top DBR, c) S3, 15% β-phase, 10.5 pairs top DBR, and d) S4, 15% β-phase/aligned, 10.5 pairs top DBR.

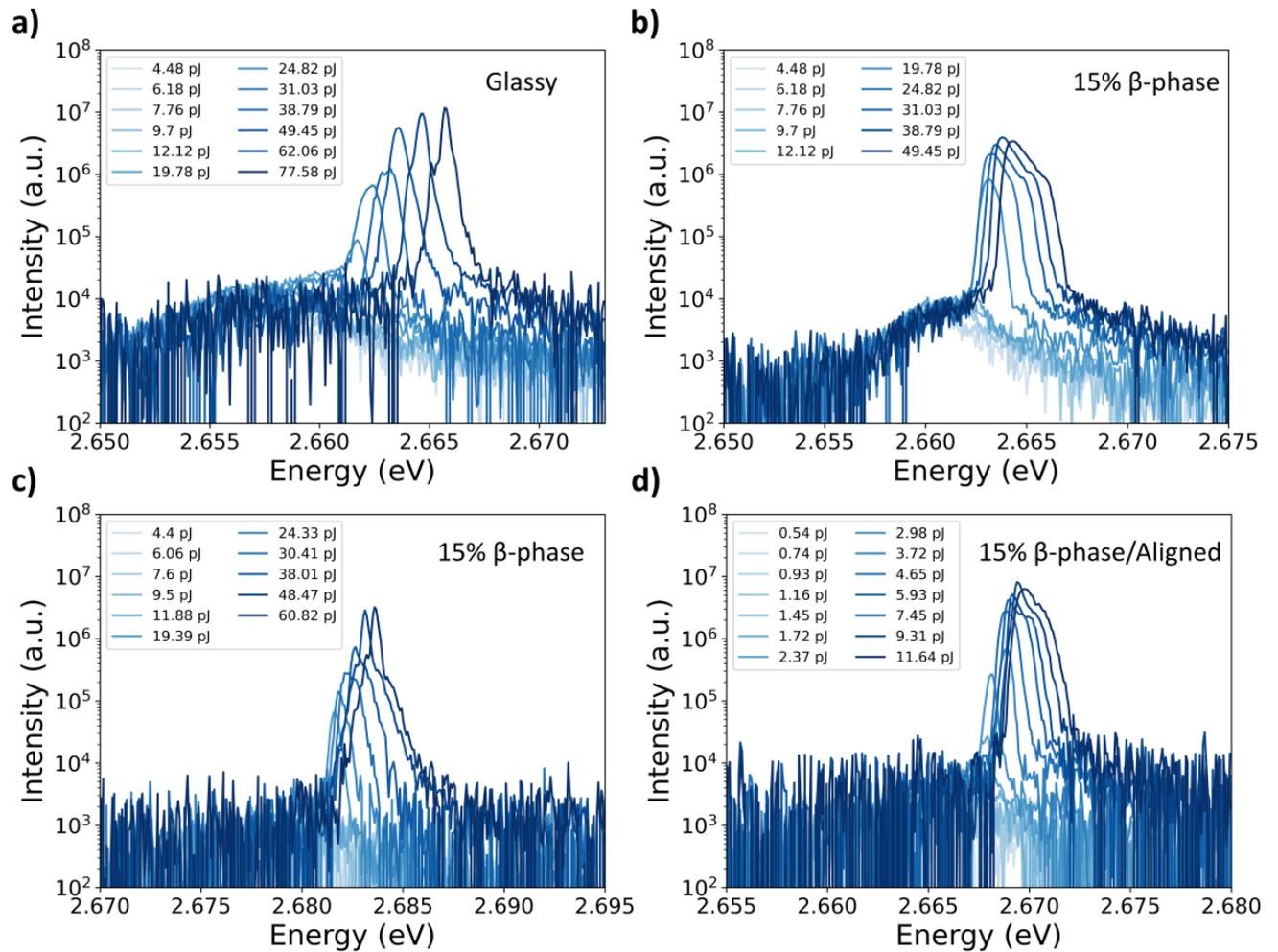

**Figure S2.** Measured transmissivity at normal incidence for samples a) S1, b) S2, c) S3 and d) S4. The dashed vertical lines indicate the low-energy edge of the stop-band for samples S1 & S2 (blue) and S3 & S4 (red). The LP was not clearly resolved in these measurements due to the high Q factors of the microcavities leading to an insufficient spectral resolution in VASE, the expected theoretical position at normal incidence ($\theta = 0°$) is represented by the black dashed line.

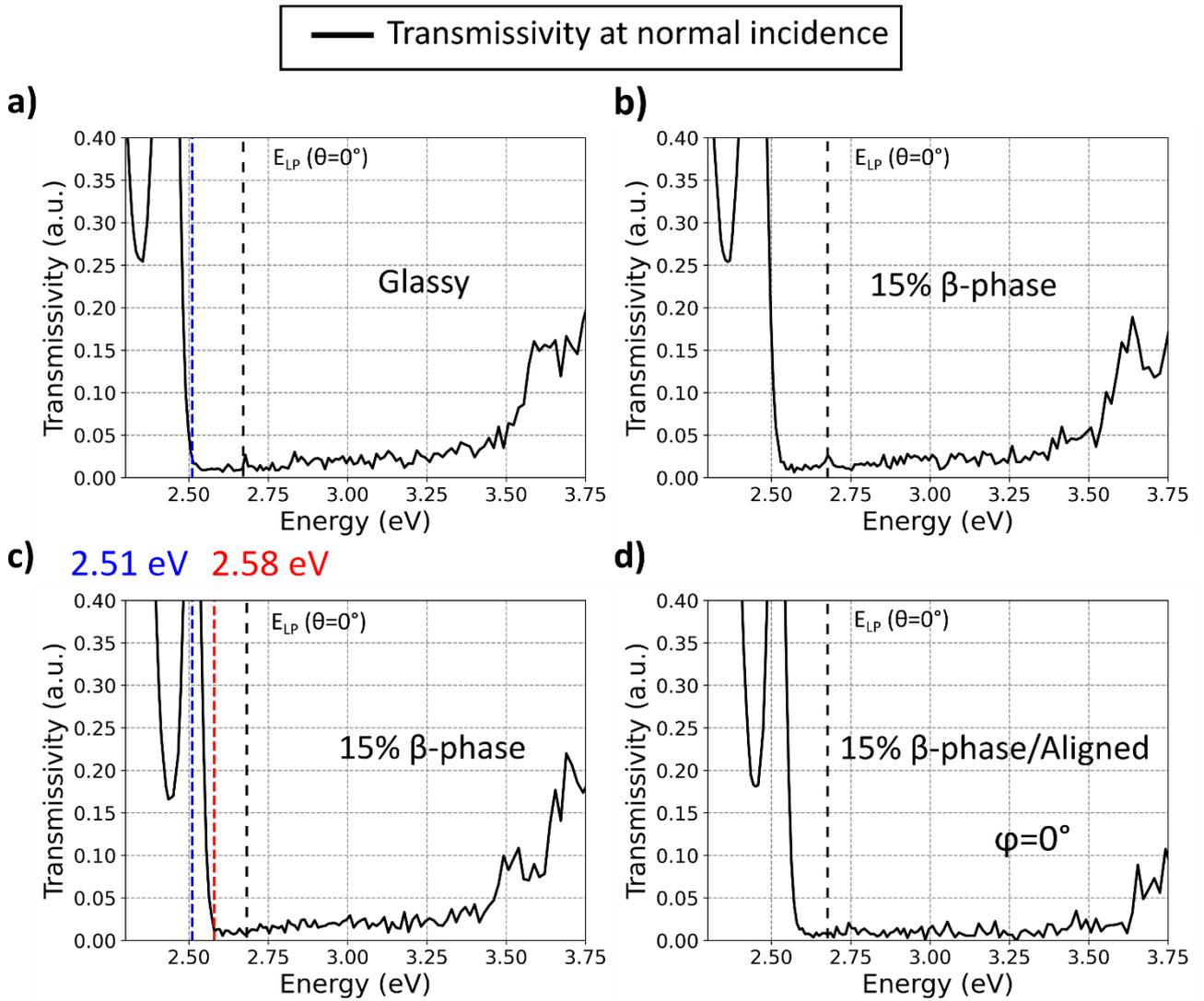

**Figure S3.** Simulation of the intra-cavity absorption of the PFO layer (blue lines), the absorption of the SD1 layer (yellow lines) and the overall reflectivity of the combined structure (black lines) for samples a) S1, b) S2, c) S3 and d) S4.

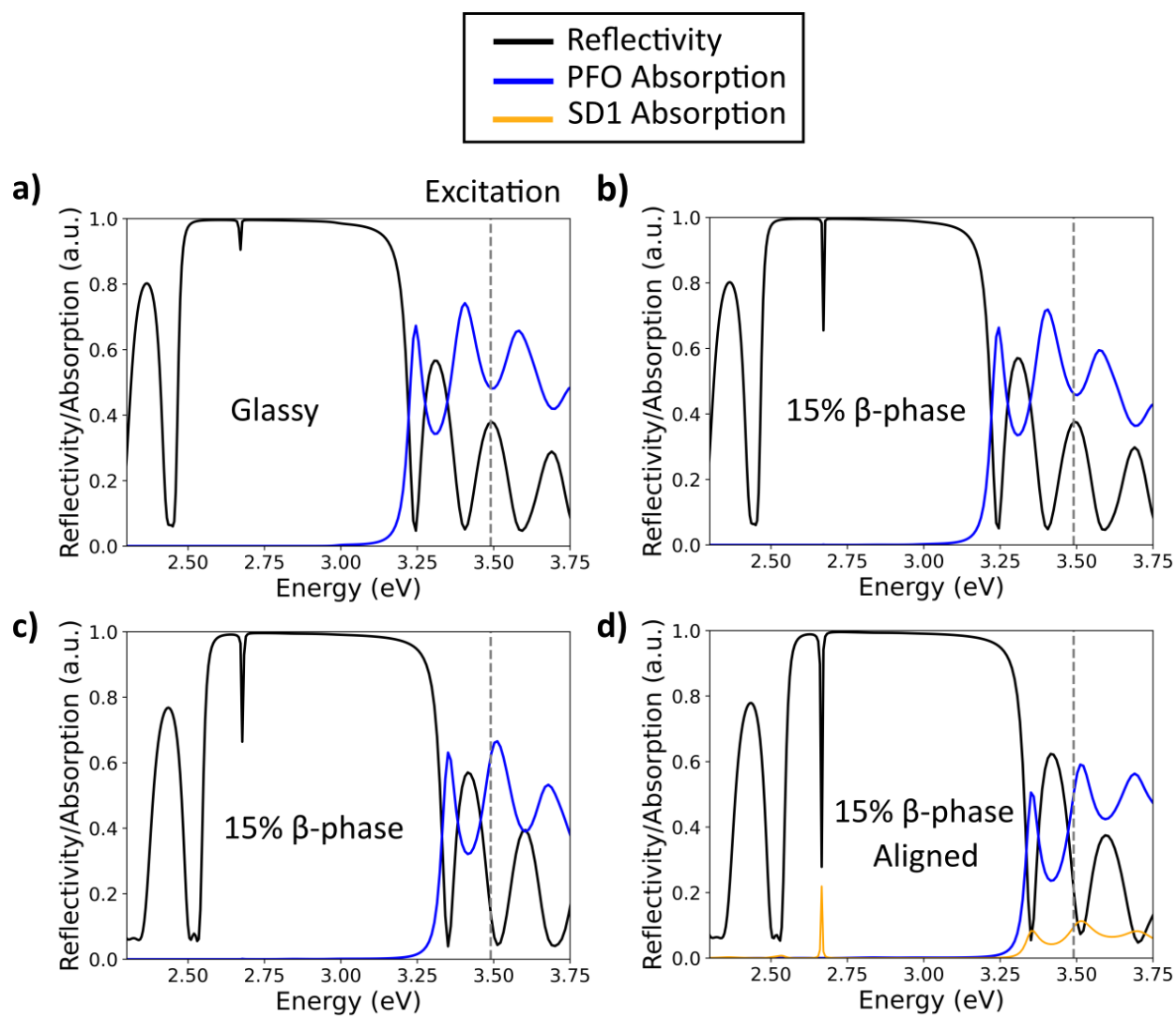

**Figure S4.** Interferometry measurements performed on samples a) S1, b) S2, c) S3, d) S4 using a Michelson interferometer in the retroreflector configuration.[1,2] The image in a) to the left of the "+" sign shows an example of the recorded image obtained by blocking one of the two arms of the interferometer, the image in a) to the right of the "+" sign and thus to the left of the "=" sign shows the mirror image obtained by freeing the blocked arm and blocking the other arm of the interferometer. Finally, the image in a) to the right of the "=" sign displays clear fringes obtained through interference of the two paths by having both arms freed, showing clear spatial coherence above threshold. b), c) and d) show the interferograms for S2, S3 and S4 with clear spatial coherence acquired above their respective thresholds.

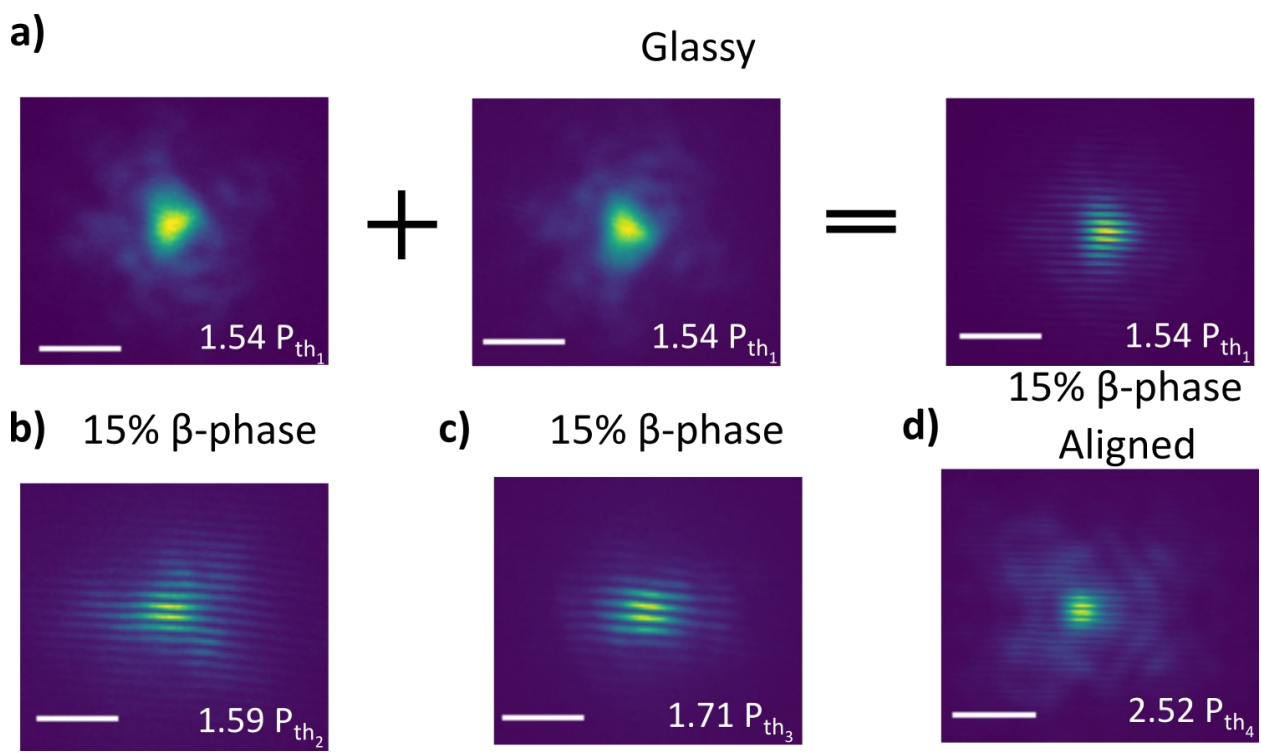